\documentclass[sn-basic, Numbered]{sn-jnl}

\usepackage{graphicx}%
\usepackage{multirow}%
\usepackage{makecell}
\usepackage{enumitem}
\usepackage{booktabs}
\usepackage{bbm}
\usepackage{microtype}
\usepackage{subfigure}
\usepackage{hyperref}

\usepackage{amsmath,amssymb,amsfonts}%
\usepackage{amsthm}%
\usepackage{mathrsfs}%
\usepackage[title]{appendix}%
\usepackage{xcolor}%
\usepackage{textcomp}%
\usepackage{manyfoot}%
\usepackage{booktabs}%
\usepackage{algorithm}%
\usepackage{algorithmicx}%
\usepackage{algpseudocode}%
\usepackage{listings}%
\usepackage{natbib}

\newcommand{\modelname}{UAlign}

\theoremstyle{thmstyleone}%
\theoremstyle{thmstyletwo}%

\theoremstyle{thmstylethree}%

\begin{document}

\title[Article Title]{UAlign: Pushing the Limit of Template-free Retrosynthesis Prediction with Unsupervised SMILES Alignment}


\author[1]{\fnm{Kaipeng} \sur{Zeng}}
\author[2]{\fnm{Bo} \sur{Yang}} 
\author[1]{\fnm{Xin} \sur{Zhao}}
\author[1]{\fnm{Yu} \sur{Zhang}}
\author[3]{\fnm{Fan} \sur{Nie}}
\author[1]{\fnm{Xiaokang} \sur{Yang}}
\author*[1]{\fnm{Yaohui} \sur{Jin}}\email{jinyh@sjtu.edu.cn}
\author*[1]{\fnm{Yanyan} \sur{Xu}}\email{yanyanxu@sjtu.edu.cn}
\affil*[1]{\orgdiv{MoE Key Laboratory of Artificial Intelligence, AI Institute}, \orgname{Shanghai Jiao Tong University}, \orgaddress{\city{Shanghai}, \postcode{200240}, \state{Shanghai}, \country{China}}}

\affil[2]{\orgdiv{Frontiers Science Center for Transformative Molecules (FSCTM), Zhangjiang Institute for Advanced Study}, \orgname{Shanghai Jiao Tong University}, \orgaddress{\city{Shanghai}, \postcode{200240}, \state{Shanghai}, \country{China}}}
\affil[3]{\orgdiv{Department of Computer Science and Engineering}, \orgname{Shanghai Jiao Tong University}, \orgaddress{\city{Shanghai}, \postcode{200240}, \state{Shanghai}, \country{China}}}


\abstract{
\textbf{Motivation} Retrosynthesis planning poses a formidable challenge in the organic chemical industry, particularly in pharmaceuticals. Single-step retrosynthesis prediction, a crucial step in the planning process, has witnessed a surge in interest in recent years due to advancements in AI for science. Various deep learning-based methods have been proposed for this task in recent years, incorporating diverse levels of additional chemical knowledge dependency. 
\vspace{2mm}\\
\textbf{Results} This paper introduces {\modelname}, a template-free graph-to-sequence pipeline for retrosynthesis prediction. By combining graph neural networks and Transformers, our method can more effectively leverage the inherent graph structure of molecules. Based on the fact that the majority of molecule structures remain unchanged during a chemical reaction, we propose a simple yet effective SMILES alignment technique to facilitate the reuse of unchanged structures for reactant generation. Extensive experiments show that our method substantially outperforms state-of-the-art template-free and semi-template-based approaches. Importantly, our template-free method achieves effectiveness comparable to, or even surpasses, established powerful template-based methods. 
\vspace{2mm}\\
\textbf{Scientific contribution} We present a novel graph-to-sequence template-free retrosynthesis prediction pipeline that overcomes the limitations of Transformer-based methods in molecular representation learning and insufficient utilization of chemical information. We propose an unsupervised learning mechanism for establishing product-atom correspondence with reactant SMILES tokens, achieving even better results than supervised SMILES alignment methods. Extensive experiments demonstrate that {\modelname} significantly outperforms state-of-the-art template-free methods and rivals or surpasses template-based approaches, with up to 5\% (top-5) and 5.4\% (top-10) increased accuracy over the strongest baseline.}




\maketitle

\section{Introduction}\label{sec1}


Retrosynthesis prediction is a crucial task in organic chemistry, aiding in finding efficient synthetic pathways from target molecules to accessible starting materials. Despite significant advancements in chemical synthesis technology, it still remains a challenge in industries like pharmaceuticals. The extensive search space and the incomplete understanding of chemical reaction mechanisms make retrosynthesis prediction difficult, even for experienced chemists. To address this issue, computer-assisted synthetic planning (CASP) has gained increasing attention in recent years, starting from the seminal work by~\citeauthor{corey1991logic}. This paper focuses on single-step retrosynthesis prediction, which is the fundamental step in CASP. It aims to predict the reactants that can lead to a given product molecule through a single reaction step.

Various deep-learning-based single-step retrosynthesis prediction methods have been proposed in recent years. These methods can be broadly classified into three groups based on their dependency on additional chemical knowledge: template-based, semi-template-based and template-free methods. \textbf{Template-based} methods~\cite{gln, Xie_Yan_Guo_Xia_Wu_Qin_2023, coley2017computer, localretro} require an extra database of reaction templates. They frame the retrosynthesis prediction as a classification or retrieval problem for reaction templates suitable for the given product molecule to be synthesized. Among these solutions, Retrosim~\cite{coley2017computer} utilizes molecular similarity to rank reaction templates; LocalRetro~\cite{localretro} and GLN~\cite{gln} use graph neural networks to model the relationship between reaction templates and molecules to predict the most suitable reaction template; RetroKNN~\cite{Xie_Yan_Guo_Xia_Wu_Qin_2023} further improves upon LocalRetro by addressing the issue of data imbalance using K-nearest neighbors (KNN). Template-based methods have strong interpretability and can accurately predict reactants. However, these methods are often unable to cover all cases and suffer from poor scalability due to limitations imposed by the template database. 

To overcome the limitations faced by template-based methods, researchers have turned to generative models. \textbf{Semi-template-based} methods incorporate chemical knowledge into generative models with the help of chemical toolkits like RDKit~\cite{landrum2013rdkit}, breaking free from the limitations imposed by reaction templates. 
The key idea of most semi-template-based methods~\cite{shi2020graph2graph, somnath2021graphretro, retroprime, Chen2023g2retro, yan2020retroxpert} is to first convert the product into synthons based on reaction center identification and then complete the synthons into reactants. Graph neural networks are commonly used for synthon prediction, followed by leaving group attachment~\cite{somnath2021graphretro, Chen2023g2retro}, conditional graph generation~\cite{shi2020graph2graph}, or SMILES generation~\cite{yan2020retroxpert} for reactant completion. Apart from all above, RetroPrime~\cite{retroprime} utilizes two independent Transformers to accomplish synthon prediction and reactant generation as separate tasks. 

Semi-template-based methods to a certain extent are more in line with chemical intuition. However, these methods increase the complexity of inference and training as they break down retrosynthesis into two subtasks. Failures in synthon prediction directly affect subsequent reactant completion and overall performance. Besides, methods based on leaving group necessitates an extra leaving group database. This requirement, akin to template-based approaches, imposes limitations on the model's scalability.

As generative models, \textbf{Template-free} methods opt to generate reactants directly from the given products. In comparison to generating graph structures, SMILES provides a way to represent molecules as strings. Taking advantage of this, most template-free methods~\cite{tetko2020state, kim2021valid, wan2022retroformer, zheng2019predicting, GTA} use Transformer models to translate between product SMILES and reactants SMILES. In particular, Graph2SMILES~\cite{tu2022permutation} replaces the Transformer encoder with a graph neural network, resulting in a permutation-invariant pipeline. There are also methods~\cite{megan,yao2024node} formulates the generation of reactants as a series of graph generation or editing operation and solve it auto-regressively. Existing template-free methods generally follows an auto-regressive generation strategy and use beam search for the generation process. Consequently, preserving a level of diversity in the resultant outputs has emerged as a critical consideration for template-free methods~\cite{vijayakumar2016diverse}. Due to the use of SMILES as input and output, most of template-free methods often overlook the rich topological and chemical bond information present in molecular graphs. Moreover, as reactants molecules need to be generated from scratch, template-free methods frequently suffer from validity issues and fail to leverage an important property of retrosynthesis prediction, i.e., the presence of many common substructures between products and reactants. 

In this paper, we focus on the template-free generative approach for retrosynthesis prediction. Existing sequence-to-sequence methods have limitations in extracting robust molecular representations. They overlook the abundance of topological information and chemical bonds, and lack the ability to utilize atom descriptors as rich as those in graph-based methods. Furthermore, template-free methods overlook the fact that the molecular graph topology remains largely unaltered from reactants to products during chemical reactions, as they generate reactants from scratch. While there are methods that attempt to solve this problem using supervised SMILES alignment, they require complex data annotation and impact model training. Given these limitations, the following question naturally arises: 
\begin{center}
    \emph{Can we effectively leverage the structural information of product molecules using a much simpler approach?}
\end{center}


To address these issues and further enhance template-free methods, we propose a novel graph-to-sequence pipeline called {\modelname}. Our approach employs a specifically designed graph neural network as an encoder, incorporating information from chemical bonds during message passing to create more powerful embeddings for the decoder. We introduce an unsupervised SMILES alignment mechanism that establishes associations between product atoms and reactant SMILES tokens which reduces the complexity of SMILES generation and enables the model to focus on learning chemical knowledge. Our model outperforms existing template-free methods by a large margin and demonstrates comparable performance against template-based methods.

\section{Methods}\label{sec11}

\begin{figure*}[htbp]
    \centering
    \includegraphics[width=\linewidth]{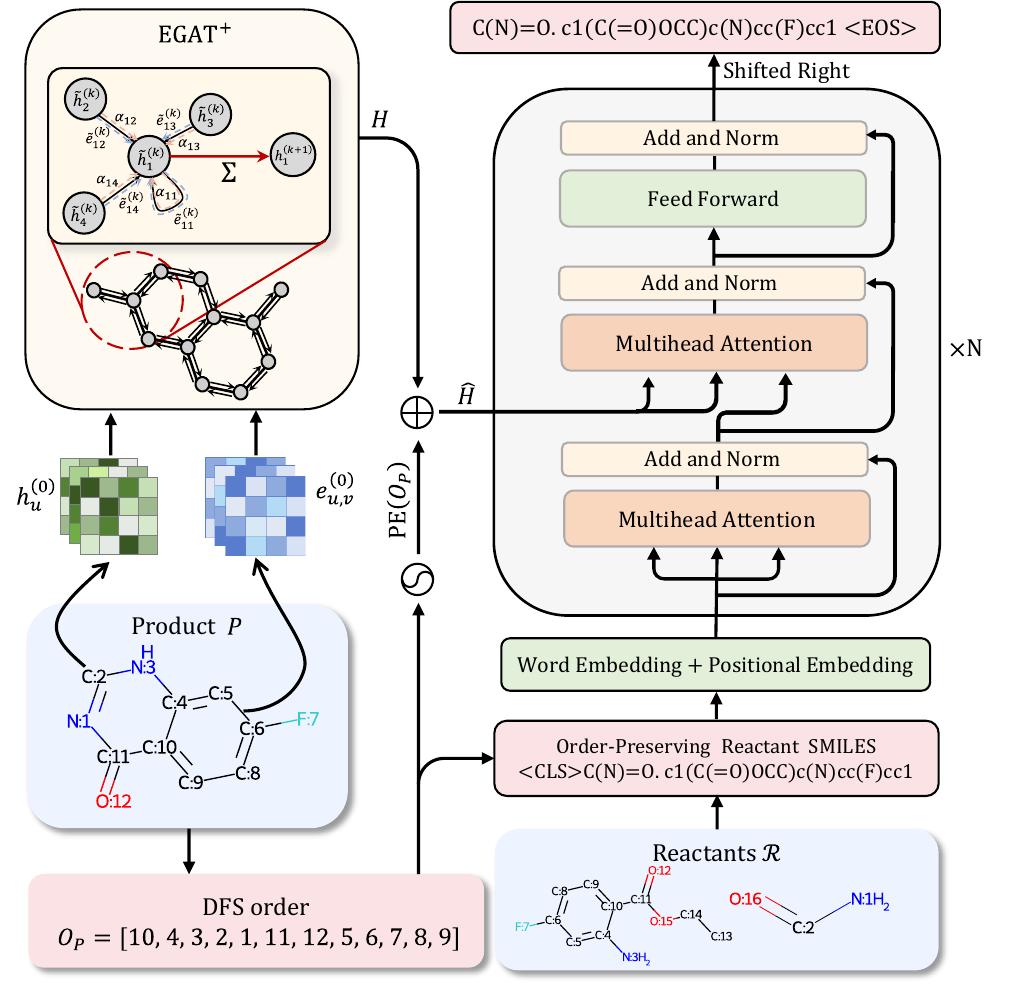}
    \caption{\textbf{Overview of {\modelname}}: Given a product molecule graph $P$ and one of its DFS order $O_P$, the graph is first fed into the graph neural network called EGAT$^+$ to obtain node features $H$. Then the positional encoding is added to $H$ according to the given DFS order $O_P$ to generate the order-aware node features $\hat{H}$. Finally the decoder takes $\hat{H}$ as input and generate the SMILES of reactants auto-regressively. }
    \label{fig:overall}
\end{figure*}

We introduce {\modelname}, a novel single-step retrosynthesis prediction model based on an encoder-decoder architecture, as demonstrated in Fig.~\ref{fig:overall}. It's a fully template-free method without any molecule editing operation using RDKit~\cite{landrum2013rdkit}. We propose a specially designed variant of Graph Attention Network, which incorporates the information of chemical bonds to enhance the capability of capturing the structural characteristics of molecules. 

\subsection{Preliminary}\label{sec: pri}
A molecule can be represented as a graph, denoted by $G=(V,E)$, where $V$ represents the atoms and $E$ represents the chemical bonds. The SMILES representation of a molecule can be obtained by performing a depth-first search (DFS) starting from any arbitrary atom in the molecule graph. 
Given a molecule graph $G=(V, E)$, we can generate multiple DFS orders and each DFS order corresponds to a SMILES representation of the graph. Denoted the set of all possible DFS orders as $\mathcal{D}(V) \subseteq \mathcal{P}(V)$, $\mathcal{P}(V)$ represents all permutations of the set of atoms $V$. 
For each DFS order $O\in\mathcal{D}(V)$, we denoted its corresponding SMILES as $Smiles(G, O)$, which lists all atoms in the order dictated by $O$. To facilitate our subsequent elaboration, we refer to the position of an atom $a$ in the order $O$ as its rank, denoted as $rank(a, O)$. The atom with the minimal rank given order $O$ is then defined as the root atom, denoted as $root(G, O)$.

\subsection{EGAT$^+$}
Chemical bonds play a significant role in determining the properties of molecules and contain valuable information. Previous studies~\cite{hu2019strategies, yang2023molerec, NEURIPS2020_94aef384} have demonstrated that incorporating edge information into graph neural networks can greatly enhance their ability to represent molecular structures. To fully leverage the information brought by chemical bonds, we propose a modified version of the Graph Attention Network (GAT)~\cite{velivckovic2018gat} called EGAT$^+$.

Our proposed model explicitly incorporates edge features, which represent the information derived from chemical bonds, into the message passing process. During each iteration of message passing, the EGAT$^+$ applies self-attention to each node and its one-hop neighbors to calculate attention coefficient according to both node features and edge features. It then aggregates the both node and edge features of these neighbors, considering the attention coefficients, to update the node features.  Denote the node feature of atom $u$ as $h^{(k)}_u$ and the edge feature between atom $u$ and $v$ as $e_{u,v}^{(k)}$ at $k$-th iteration of message passing. In math, the message passing mechanism can be written as 
\begin{equation}\label{gat}
\begin{aligned}    
    \tilde{e}_{u,v}^{(k)} &= \mathrm{FFN}^{(k)}_e (e_{u,v} ^{(k)}),\\
    \tilde{h}_{u}^{(k)}& = \mathrm{FFN}^{(k)}_n (h_u^{(k)}),\\
    c_{u,v} &= \mathbf{a}^T [\tilde{h}_u^{(k)} \Vert \tilde{h}_v^{(k)}\Vert \tilde{e}_{u,v}^{(k)}],\\
    \alpha_{u,v} &= \frac{\exp(\mathrm{LeakyReLU}(c_{u,v}))}{\sum_{v'\in \mathcal{N}(u)\cup \{u\}} \exp(\mathrm{LeakyReLU}(c_{u,v'}))},\\
    h^{(k+1)}_u &= \sum_{v\in \mathcal{N}(u) \cup \{u\}} \alpha_{u,v} \left(\tilde{h}_{u}^{(k)} + \tilde{e}_{u,v} ^ {(k)}\right),\\
    e^{(k+1)}_{u,v} &= \mathrm{FFN}_m^{(k)}([h_{u}^{(k+1)}\Vert h_{v}^{(k+1)} \Vert e^{(k)}_{u,v}]), u\neq v,
\end{aligned}
\end{equation}
where $\mathrm{FFN}_m^{(k)}$, $\mathrm{FFN}^{(k)}_e$ and $\mathrm{FFN}^{(k)}_n$ are three different feed forward networks, $\mathbf{a}$ is a learnable parameter, $\mathcal{N}(u)$ denotes the one-hop neighbors of node $u$ and $\Vert$ denotes the concatenation operation.
Since there are no chemical bonds with the same beginning and ending atoms, the $e^{(k)}_{u, u}$ is also set as a learnable parameter shared among all atoms. The residual connection and layer normalization~\cite{2016Layer} are applied to prevent over-smoothing while enlarging the receptive field of the model~\cite{wu2022nodeformer}. 

The initial node features $h_u^{(0)}$ and edge features $e_{u,v}^{(0)}$ are determined via several chemical property descriptors, whose details are shown in Supplementary Sec. 6. After $K$ iterations of message passing, we can obtain the encoded features $h^{(K)}_u$ of all atoms and make up the output $H\in \mathbb{R}^{V_P \times d}$ of the encoder, where $d$ denotes the embedding size.

\subsection{SMILES Alignment}\label{sec: smalign}
For single-step retrosynthesis prediction, a significant proportion of structures are shared between product molecules and reactant molecules~\cite{wan2022retroformer, zhong2022rootalign}. However, SMILES-based methods often have to generate the reactant SMILES from scratch, even if most of the structures of reactants are the same as those of the products. This results in the underutilization of input information and becomes the bottleneck of template-free retrosynthesis prediction methods. There are methods~\cite{wan2022retroformer,GTA} addressing this issue through supervised SMILES alignment, which involves adding supervised information to establish the correspondence between input and output tokens through cross-attention over the input and predicted tokens. This supervised training approach not only requires complex data annotation algorithms but also limits the diversity of the model's attention map, thereby further affecting the model's performance. To address the above-mentioned issues, we propose the unsupervised SMILES alignment method as follows.

Assuming we can identify the location of each product atom in the reactants' SMILES and provide it to the model, a natural correspondence can be established between the input and output atoms. However, during the inference process, revealing this information would lead to label leakage, which is not permitted. Therefore, we propose the following modification: when providing an order of product atoms, we expect the model to generate atom tokens in the reactants' SMILES in this given order as closely as possible. By doing so, we can establish a correspondence between the product atoms and the reactants' SMILES tokens using unsupervised methods without leaking any labels. We refer to this type of reactants' SMILES, which aims to preserve the given order of atom tokens as much as possible, as order-preserving reactant SMILES. Note that SMILES represents atoms in a molecule according to a certain DFS order, the provided order should also be a DFS order for the product molecule.

The generation of order-preserving reactant SMILES will be introduced as follows. Given the product molecule $P=(V_P,E_P)$ with a DFS order $O_P\in \mathcal{D}(V_P)$ and the corresponding set of reactant molecules $\mathcal{R}=\{R_1, R_2,\ldots R_l\}$, for each reactant $R=(V_R, E_R)\in \mathcal{R}$, we can find a depth-first order $O_R \in \mathcal{D}(V_R)$ that has a nearly consistent atomic appearance sequence with $O_P$ as the product and reactants have similar structures. For convenience, we name such a order as $O_P$-corresponding order of $R$ and denote it as $CO(R, O_P)$. Mathematically, it's defined as 
\begin{equation}\label{eq:co}
    CO(R, O_P) = \arg\min_{o\in \mathcal{D}(R)} \sum_{i\in O_P\cap o} \sum_{j\in O_P\cap o} inv(i, j, O_P, o),
\end{equation}
where the value of $inv(i, j, O_P, o)$ equals $1$ if and only if  $rank(i, O_P) < rank(j,O_P)$ and $rank(i,o) > rank(j, o)$, and equals $0$ otherwise. We sort the reactants $\mathcal{R}$ according to  $rank(root(R, CO(R, O_P)), O_P)$ in ascending order. Then we generate SMILES for each reactant molecule using its $O_P$-corresponding order and join them together using ``.'' to obtain order-preserving reactant SMILES. 

For further discussion, we denote the order-preserving reactant SMILES given the reactant molecules $\mathcal{R}$ and a DFS order $O$ of product as $OPSmiles(\mathcal{R}, O)$. An example of the process to generate order-preserving reactants SMILES is shown in Fig.~\ref{fig:orsmiles-exm}.  The detailed implementations are presented in Supplementary Sec. 5.1.

\begin{figure}[t]
    \centering
    \includegraphics[width=\linewidth]{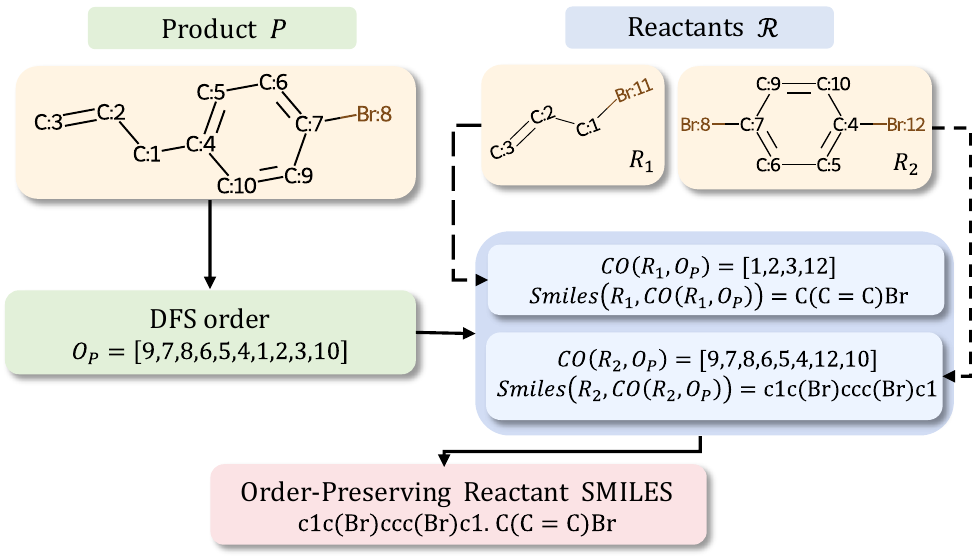}
    \caption{An example of the process to generate order-preserving reactants SMILES. The atom mapping numbers shown on the figure are included only for clearer explanation and  will be removed in our implementation to prevent any label leakage.}
    \label{fig:orsmiles-exm}
\end{figure}
\subsection{Decoder}
The decoder takes the node features $H\in \mathbb{R}^{V_P \times d}$ that are generated from the encoder, as well as the given DFS order $O_P$ for the product molecule graph as input. We use the vanilla Transformer decoder~\cite{vaswani2017attention} as our decoder. 
As mentioned in Sec.~\ref{sec: smalign}, the order information of product atoms are required for SMILES alignment. However, the Transformer decoder is permutation-invariant to memory~\cite{tu2022permutation, pmlr-v97-lee19d}, meaning it is not sensitive to the order of the features from encoder. This implies that directly performing cross-attention over $H$ may not effectively capture the relationship between product atoms and reactant SMILES tokens. To address this problem, we introduce position encoding to the node features based on the rank of each atom in the given DFS order $O_P$ to generate order-aware node features $\hat{H}$. Then given an input embedding sequence $Z\in \mathbb{R}^{m\times d}$, the Transformer decoder layer utilizes the order-aware node features as keys and values in all the cross-attention layers. This process ultimately generates the decoded embeddings $\hat{Z}\in \mathbb{R}^{m\times d}$. These embeddings are then fed into feed-forward layers $\mathrm{FFN_1}: \mathbb{R}^{d}\rightarrow \mathbb{R}^T$ to predict the tokens $\hat{T}$ that should be generated. In summary, the decoder can be mathematically expressed as
\begin{equation}\label{eq:PE}
    \begin{aligned}
        \hat{H} &= H + \mathrm{PE}(O_P),\\
        \hat{Z}&=\mathrm{TransformerDecoder}(Z, \hat{H}),\\
        \hat{T} &= \mathrm{FFN_1}(\hat{Z}).
    \end{aligned}
\end{equation}

\subsection{Two stage training}

There is a significant distribution shift between graphs and SMILES representations. Moreover, our model is specifically designed to generate non-canonical SMILES, which may contain more complex patterns compared to canonical SMILES. To achieve this goal, we propose a two-stage training strategy in this paper. The first stage aims to align the distributions between two distinct modalities: SMILES and molecular graphs, while enabling the model to learn the patterns of non-canonical SMILES. Given a molecule graph $M$ and one of its possible DFS orders $O_M$, the training task is to translate the graph into the corresponding SMILES representation based on the given order $O_M$. In detail, this is reached by training the model to generate $Smiles(M, O_M)$ given molecule $M$ and DFS order $O_M$.

Once the first stage training converges, we proceed to the second stage, which focuses on retrosynthesis prediction. In this stage, the model is trained using the order-preserving reactant SMILES as targets. Given a product molecule graph $P$, a possible DFS order $O_P$, and a set of reactants $\mathcal{R}$, the model is expected to generate $OPSmiles(\mathcal{R}, O_P)$.


\subsection{Data Augmentation}
Different from those Transformer-based methods~\cite{wan2022retroformer, tetko2020state, GTA} taking SMILES as input and canonical SMILES as target, our method takes a graph as input and is trained with non-canonical SMILES. That means the previous SMILES augmentation tricks are not suitable for us. Similar to~\cite{wan2022retroformer}, we choose to augment the training data on-the-fly.

For the first stage, at each iteration, for each molecule $M=(V_M, E_M)$, we have a 50\% chance of using a random DFS order $O_M$ as the input for the model, and using the corresponding $Smiles(M, O_M)$ as the training target. For the other 50\%, we randomly select another molecule $M'=(V_{M'}, E_{M'})$ from the dataset to form a new molecular graph $\tilde{M}=(V_M \cup V_{M'}, E_M \cup E_{M'})$, and find the DFS order $O_{\tilde{M}}$ that can generate canonical SMILES for $\tilde{M}$. $\tilde{M}$ and $O_{\tilde{M}}$ are then fed into the model and the target is set as the canonical SMILES of $\tilde{M}$. Such an augmentation method enables the model to output the atom tokens according to the given DFS order and be aware of different components within a graph.

For the second stage, at each iteration, for each product molecule $P=(V_P, E_P)$, we have a 50\% probability of using a random DFS order as input, and for the remaining part, we use the DFS order capable of producing canonical SMILES for product as input. The target used for training is the order-preserving reactant SMILES generated based on the input DFS order. This data augmentation method allows the model to focus more on the DFS order for canonical product SMILES while also noticing the correspondence between product atoms and the output SMILES tokens.

\subsection{Loss}
Both stage of training can considered as a kind of translation between graphs and SMILES, thus we use the loss widely used for auto-regressive language generation models for training. Denote the training target as $T=\{t_1, t_2, \ldots, t_n\}$ and the output of the model $\hat{T}=\{\hat{t}_1, \hat{t}_2,\ldots,\hat{t}_n\}$, the loss can be written as 
\begin{equation}
    \mathcal{L}=\sum_{i=1}^n l_{cls}(\hat{t}_i, t_i),
\end{equation}
where $l_{cls}(\cdot)$ is the classification loss.

\section{Results and Discussion}\label{sec2}


In this section, we conduct extensive experiments to make a comprehensive evaluation of our proposed {\modelname}. 

\subsection{Evaluation Protocol}\label{sec: baselines}

\textbf{Benchmark Datasets.} We adopt three datasets for evaluation: (1) \textbf{USPTO-50K} consists of 50,016 atom-mapped reactions grouped into 10 different classes; (2) \textbf{USPTO-FULL }comprises 1,013,118 atom-mapped reactions without any reaction class information. (3) \textbf{USPTO-MIT} consists of 479,035 atom-mapped reactions without any reaction class information. To ensure a fair comparison, we adopt the same training/validation/test splits as those in a previous study~\cite{gln} for USPTO-50K and USPTO-FULL datasets. The training/validation/test splits is aligned with the previous study~\cite{jin2017predicting}. The detailed data processing procedure and the statistical information of the processed dataset are presented in Supplementary Sec. 2.

\textbf{Metrics.} We utilize the following three evaluation metrics for evaluation: \textbf{top-$k$ accuracy}, \textbf{top-$k$ SMILES validity}, and \textbf{top-$k$ round-trip accuracy}. The detailed definitions for three metrics are provided in Supplementary Sec. 3.

\subsection{Performance Comparison}

\begin{table}[htbp]
\centering
\caption{Top-$k$ accuracy for retrosynthesis prediction on USPTO-50K. * indicates the model with SMILES augmentation. For comparison purpose, the Aug. Transformer is evaluated without the test augmentation. Best performance of each model type is in \textbf{bold}.}\label{tab:overall}
\begin{tabular}{lccccccccc}
\toprule
\multicolumn{1}{c}{\multirow{3}{*}{\textbf{Model}}} & \multicolumn{9}{c}{\textbf{Top-$k$ accuracy (\%)}} \\ 
\cmidrule{2-10} 
& \multicolumn{4}{c}{\textbf{Reaction class known}} & \textbf{} & \multicolumn{4}{c}{\textbf{Reaction class unknown}} \\ 
\cmidrule{2-5} \cmidrule{7-10}
& 1 & 3 & 5 & 10 & & 1 & 3 & 5 & 10\\
\hline
\textbf{Template-Based}\\ 
\hline
RetroSim~\cite{coley2017computer}& 52.9&73.8&81.2&88.1&& 37.3&54.7&63.3&74.1\\
NeuralSym~\cite{segler2017neural}&55.3&76.0&81.4&85.1&&44.4&65.3&72.4&78.9\\
GLN~\cite{gln} & 64.2 & 79.1 & 85.2 & 90.0 & & 52.5 & 69.0 & 75.6 & 83.7 \\
LocalRetro~\cite{localretro} & 63.9 & 86.8 & 92.4  & 96.3 & & 53.4 & 77.5 & 85.9 & 92.4\\ 
RetroKNN \cite{Xie_Yan_Guo_Xia_Wu_Qin_2023} & \textbf{66.7} & \textbf{88.2} & \textbf{93.6} & \textbf{96.6}&  &\textbf{57.2}& \textbf{78.9}& \textbf{86.4}& \textbf{92.7}\\
\hline
\textbf{Semi-Template-Based}\\ 
\hline
RetroXpert*~\cite{yan2020retroxpert} & 62.1 & 75.8 & 78.5 & 80.9 & & 50.4 & 61.1 & 62.3 & 63.4 \\
G2G~\cite{shi2020graph2graph} & 61.0 & 81.3 & {86.0} & {88.7} & & 48.9 & 67.6 & 72.5 & 75.5 \\
GraphRetro~\cite{somnath2021graphretro} & 63.9 & 81.5 & \textbf{85.2} & \textbf{88.1} & & \textbf{53.7} & 68.3 & 72.2 & 75.5\\
RetroPrime*~\cite{retroprime} & \textbf{64.8} & \textbf{81.6} & 85.0 & 86.9 & & 51.4 & \textbf{70.8} & \textbf{74.0} & \textbf{76.1}\\ 
\hline
\textbf{Template-Free} \\ 
\hline
Transformer~\cite{vaswani2017attention} & 57.1 & 71.5 & 75.0 & 77.7 & & 42.4 & 58.6 & 63.8 & 67.7\\
SCROP~\cite{zheng2019predicting} & 59.0 & 74.8 & 78.1 & 81.1 & & 43.7 & 60.0 & 65.2 & 68.7\\
Liu's Seq2seq~\cite{liu2017retrosynthetic} &- &-&-&-&&37.4 &52.4 &57.0 &61.7\\
Tied Transformer~\cite{kim2021valid} & - & - & - & - & & 47.1 & 67.1 & 73.1 & 76.3 \\
Aug. Transformer* \cite{tetko2020state}& - & - & - & - & & 48.3 & - & 73.4 & 77.4 \\
MEGAN~\cite{megan} & 60.7 & 82.0 & 87.5 & 91.6 &  & 48.1 & 70.7  & 78.4  & 86.1\\ 
GTA* \cite{GTA} & - & - & - & - & & 51.1 & 67.6 & 74.8 & 81.6\\
Graph2SMILES~\cite{tu2022permutation} & - & - & - & - & & {52.9} & 66.5 & 70.0 & 72.9  \\
Retroformer*~\cite{wan2022retroformer} & 64.0 & 82.5 & 86.7 & 90.2 & & 53.2 & 71.1 & 76.6 & 82.1 \\ 
RetroBridge~\cite{igashov2023retrobridge}& -&-&-&-&&50.8 & 74.1&80.6& 85.6\\
Ours* & \textbf{66.2}&	\textbf{86.9}&	\textbf{91.9}	&\textbf{95.1} & & \textbf{53.6} & \textbf{77.6} & \textbf{84.6} & \textbf{90.3} \\
\Xhline{1pt}
\end{tabular}
\end{table}
\begin{table}[htbp]
    \centering
    \caption{Top-$k$ accuracy for retrosynthesis prediction on USPTO-MIT. Best performance of each model type is in \textbf{bold}.}
    \begin{tabular}{llcccc}
    \toprule
       \multicolumn{1}{c}{\multirow{2}{*}{\textbf{Model Type}}} &\multicolumn{1}{c}{\multirow{2}{*}{\textbf{Model}}} &  \multicolumn{4}{c}{\textbf{Top-$k$ accuracy (\%)}} \\
        \cmidrule{3-6}
        & &  1 & 3 & 5&10\\
        \midrule
        \multirow{2}{*}{template-based}     &NeuralSym~\cite{segler2017neural} & 47.8 &67.6 &74.1 &80.2\\ 
    & LocalRetro~\cite{localretro}&\textbf{54.1}& \textbf{73.7} &\textbf{79.4}& \textbf{84.4} \\
    \hline 
     \multirow{4}{*}{template-free}& Liu's Seq2seq~\cite{liu2017retrosynthetic} &46.9& 61.6 &66.3 &70.8\\
     &AutoSynRoute~\cite{lin2020automatic}& 54.1& 71.8& 76.9& 81.8\\
    &RetroTRAE~\cite{ucak2022retrosynthetic}& 58.3&-&-&- \\
    &Ours & \textbf{59.9} & \textbf{76.9} & \textbf{82.0} & \textbf{86.4}\\
    \bottomrule
    \end{tabular}
    \label{result: mit}
\end{table}
\begin{table}[htbp]
    \centering
    \caption{Top-$k$ accuracy for retrosynthesis prediction on USPTO-FULL. * indicates the model with SMILES augmentation. Best performance of each model type is in \textbf{bold}.}
    \begin{tabular}{llcccc}
    \toprule
       \multicolumn{1}{c}{\multirow{2}{*}{\textbf{Model Type}}} &\multicolumn{1}{c}{\multirow{2}{*}{\textbf{Model}}} &  \multicolumn{4}{c}{\textbf{Top-$k$ accuracy (\%)}} \\
        \cmidrule{3-6}
        & &  1 & 3 & 5&10\\
        \midrule
        \multirow{4}{*}{template-based} &RetroSim~\cite{coley2017computer}& 32.8 & -&-&56.1\\
        &NeuralSym~\cite{segler2017neural} & 35.8 & -&-&60.8\\ 
        &GLN~\cite{gln}  &\textbf{39.3}  & -&-&\textbf{63.7}  \\
    & LocalRetro~\cite{localretro}&39.1 &  \textbf{53.3}&\textbf{58.4}&63.7 \\
    \hline
    semi-template-based & RetroPrime*~\cite{retroprime} & \textbf{44.1} &\textbf{59.1}& \textbf{62.8} &\textbf{68.5} \\
    \hline
     \multirow{5}{*}{template-free}&MEGAN~\cite{megan} & 33.6& - & - &63.9\\
     &Aug. Transformer*~\cite{tetko2020state} & 46.2 &- &-& 73.3\\
    &Graph2SMILES~\cite{tu2022permutation}  & 45.7&-&-&63.4 \\
    &GTA*~\cite{GTA}& 46.6&-&-&70.4\\
    &Ours* & \textbf{50.4}	&\textbf{66.1} &	\textbf{71.3}	&\textbf{76.2}\\
    \bottomrule
    \end{tabular}
    \label{result: full}
\end{table}

\begin{table}[htbp]
\centering
    \caption{Top-$k$ SMILES validity for retrosynthesis prediction on USPTO-50K with reaction class unknown. }
    \label{tab:validity}
    \begin{tabular}{lcccc}
    \toprule
    \multicolumn{1}{c}{\multirow{2}{*}{\textbf{Model}}} & \multicolumn{4}{c}{\textbf{Top-$k$ validity (\%)}} \\ 
    \cmidrule{2-5} 
     & \multicolumn{1}{c}{1}  & \multicolumn{1}{c}{3} & \multicolumn{1}{c}{5} & \multicolumn{1}{c}{10} \\ 
    \hline
    Transformer~\cite{vaswani2017attention} & {97.2} & {87.9} & {82.4} & {73.1} \\
    Graph2SMILES~\cite{tu2022permutation}  & 99.4 & 90.9 & 84.9 & 74.9 \\
    RetroPrime~\cite{retroprime} & 98.9 & 98.2 & 97.1 & 92.5 \\
    Retroformer~\cite{wan2022retroformer} & {99.2} & {98.5} & {97.4} & 96.7 \\ 
    Ours &  \textbf{99.8} & \textbf{99.7}& \textbf{99.3}& \textbf{98.2}\\
    \Xhline{1pt}
    \end{tabular}
\end{table}
\begin{table}[htbp]
    \centering
    \caption{Top-$k$ round-trip accuracy for retrosynthesis prediction on USPTO-50K with reaction class unknown.}
    \label{tab:round_trip}
    \begin{tabular}{lcccc}
    \toprule
    \multicolumn{1}{c}{\multirow{2}{*}{\textbf{Model}}} & \multicolumn{4}{c}{\textbf{Top-$k$ round-trip acc. (\%)}} \\ \cmidrule{2-5} 
     & 1 & 3 & 5 & 10 \\ \hline
     GraphRetro~\cite{somnath2021graphretro} & 80.5 & 73.3& 68.3&59.3\\
    Transformer~\cite{vaswani2017attention} & 71.9 & 54.7 & 46.2 & 35.6 \\
    Graph2SMILES~\cite{tu2022permutation} & 76.7 & 56.0 & 46.4 & 34.9 \\
    RetroPrime~\cite{retroprime} & 79.6 & 59.6 & 50.3 & 40.4 \\
    Retroformer~\cite{wan2022retroformer} & {78.9} & 72.0 & {67.1} & {57.2} \\ 
    Ours &\textbf{80.9} & \textbf{74.0}& 	\textbf{69.0}& \textbf{60.2}\\
    \Xhline{1pt}
    \end{tabular}
\end{table}

\textbf{Top-$k$ Accuracy.}
We compare our model with existing single-step retrosynthesis prediction in terms of top-$k$ accuracy on all the datasets. The results are summarized in Table~\ref{tab:overall}, Table~\ref{result: mit} and Table~\ref{result: full}. On the USPTO-50K dataset, our model achieves a top-3 accuracy of 77.6\%, top-5 accuracy of 84.6\% and top-10 accuracy of 90.3\% under the reaction class unknown setting, surpassing the SOTA template-free method by 3.5\%, 4.0\% and 4.7\% respectively. And with reaction class given on USPTO-50K dataset, our model achieves the top-1 accuracy of 66.2\%, top-5 accuracy of 91.9\% and top-10 accuracy of 95.1\%, which exceeds the SOTA template-free method by 2.2\%, 4.4\% and 4.9\% respectively. Moreover, our model  outperforms all the semi-template-based methods with a noticeable margin. It's also encouraging to see that our method, as a template-free method, achieves competitive or even superior performance against the powerful template-based methods such as LocalRetro under both settings of USPTO-50K dataset. On UPSTO-MIT dataset, our model achieves the top-1 accuracy of 59.9\% and top-10 accuracy of 86.4\%, which even outperforms the existing template-based SOTA method LocalRetro significantly. Additionally, our models achieved a top-1 accuracy of 50.4\% on the USPTO-FULL dataset, which exceed that of the current SOTA model GTA by 3.8\%. These findings sufficiently demonstrate the effectiveness of our method. The contribution of each proposed module will be further validated in Sec.~\ref{sec: abla}.

It is noteworthy that while template-based approaches have achieved remarkable performance on the USPTO-50K dataset, their reliance on external template libraries has emerged as a constraint as datasets grow in scale and complexity. This dependency leads to a substantial degradation in model performance.  In contrast, template-free methods have demonstrated superior versatility and adaptability, qualities that render them especially appropriate for managing large-scale and intricate datasets.

\textbf{Top-$k$ SMILES Validity.} We use vanilla Transformer, RetroPrime, Retroformer and Graph2SMILES as robust baselines to compare the validity of SMILES in our study. SMILES generation models for retrosynthesis tasks often encounter challenges with maintaining SMILES validity. We do not take the methods based on templates or molecule editing as baseline here because the validity of generated SMILES can be guaranteed by the templates or chemical toolkits. Unlike graph-based models, SMILES-based methods need to ensure that the generated content adheres to the parsing rules of SMILES, without leveraging chemical tools such as RDKit. Consequently, SMILES-based approaches are more susceptible to generating invalid SMILES compared to graph-based models. As shown in Table~\ref{tab:validity}, our model demonstrates superior top-1 and top-5 molecule validity compared to other models, even without employing canonical SMILES as our training objective. This improvement could be attributed to the proposed two-stage training strategy and data augmentation, which assist the model in capturing various SMILES patterns effectively.

\textbf{Top-$k$ Round-Trip Accuracy.} To assess the accuracy of our predicted synthesis plans, we utilize the Molecule Transformer~\cite{schwaller2019molecular} as the benchmark reaction prediction model and calculate the top-$k$ round-trip accuracy. We take RetroPrime,  Retroformer and Graph2SMILES as our strongv SMILES-based baselines. We also use take graph-based method GraphRetro into comparison. The results are presented in Table~\ref{tab:round_trip}. The results clearly indicate that our model outperforms all SMILES-based baselines by a considerable extent and even exceeds the well-established graph-based method, GraphRetro. This underscores the efficacy of our unsupervised SMILES alignment mechanism, which enables the model to efficiently leverage substructures from product molecules to construct reactants. This mechanism allows the model to focus more intently on learning reaction mechanisms, thereby yielding more plausible predictive outcomes. In summary, our model has exhibited a robust capacity for generating coherent and efficacious synthesis pathways, specifically tailored for advanced downstream applications such as multi-step retrosynthesis planning.

\subsection{Ablation Study}\label{sec: abla}

\begin{sidewaystable}[htbp]
    \centering
    \caption{Effects of different modules on retrosynthesis performance in reaction class unknown setting of USPTO-50K dataset. Best performance is in \textbf{bold}.}
    \label{tab:ablation}
    \begin{tabular}{l cccc c cccc c cccc}
    \toprule
       \multicolumn{1}{c}{\multirow{2}{*}{\textbf{Method}}}  & \multicolumn{4}{c}{\textbf{Top-$k$ acc. (\%)}}& & \multicolumn{4}{c}{\textbf{Top-$k$ round-trip acc. (\%)}} &  &\multicolumn{4}{c}{\textbf{Top-$k$ validity (\%)}}  \\
       \cmidrule{2-5} \cmidrule{7-10} \cmidrule{12-15}
         & 1& 3 & 5 & 10 & & 1 & 3 & 5 & 10 & & 1 & 3 & 5 & 10\\
        \midrule
        {\modelname} (Full Version)& \textbf{53.6} & \textbf{77.6} & \textbf{84.6} & \textbf{90.3} & & \textbf{80.9} & \textbf{74.0}& 	\textbf{69.0}& \textbf{60.2} & & \textbf{99.8} & \textbf{99.7}& \textbf{99.3}& \textbf{98.2} \\
        $-$ Two Stage Training & 53.1&	75.6	&82.8	&88.8 & & 80.4&	72.3&	66.9&	57.6 & &99.2&	97.7&	96.1 &	92.2\\
        $-$ Data Augmentation & 48.5&71.8&79.2&85.9 & & 78.2&70.5&	65.1&56.0 & & 99.1&98.4&97.7&	95.6 \\
        $-$ SMILES Alignment & 45.5	&65.2	&71.4&	77.5& &	70.3&57.9&	51.0&40.1 && 97.8&95.3&	93.6&89.2\\
        \midrule
        Transformer~\cite{vaswani2017attention} & 42.4 & 58.6 & 63.8 & 67.7 & & 71.9 & 54.7 & 46.2 & 35.6  && {97.2} & {87.9} & {82.4} & {73.1}\\
        \bottomrule
    \end{tabular}
\end{sidewaystable}
We investigate the effects of different components in our proposed pipelines. The result is summarized in Table~\ref{tab:ablation}.

\textbf{Two Stage Training.} We eliminate the initial training phase and directly train the model for the retrosynthesis prediction task. As indicated in Table~\ref{tab:ablation}, the two-stage training strategy has consistently led to enhancements in all evaluated metrics. This observation implies that the two-stage training strategy effectively enables  the model to adeptly learn the intricacies of molecular SMILES representations, thereby yielding higher quality and more plausible retrosynthetic analysis outcomes. 

\textbf{Data Augmentation.} We remove the data augmentation during the second training stage, which means training solely using the DFS order that can generate canonical SMILES. Table~\ref{tab:ablation} demonstrates a significant decline in model performance across all metrics. This clearly demonstrates that our data augmentation significantly improves the model's performance.

\textbf{SMILES Alignment.} In the training process, we remove all operations related to SMILES alignment. This includes the removal of the position encoding in Eq.~\ref{eq:PE}, where the features $H$ directly served as the input memory for the Transformer decoder. Since we eliminate the input related to the DFS order, the model was no longer trained using order-preserving reactants SMILES as the target but instead switched to canonical SMILES for product. Additionally, in this set of experiments, we remove the first training stage, which aligns the graph and SMILES modalities as the model architecture changes. The results are reported in Table~\ref{tab:ablation}, and they show a significant decline in performance compared to our full version, indicating that the proposed SMILES alignment algorithm is crucial for achieving excellent performance.

It is worth noting that even without data augmentation, two-stage training and SMILES alignment, our model still outperforms the vanilla Transformer by a large margin in terms of all metrics reported in the last line of Table~\ref{tab:ablation}. This indicates that graph-based molecular representation learning still has advantages over SMILES-based approaches, and our proposed EGAT$^+$ can extract effective molecular representations for downstream usage.
\subsection{Case Study (Visualization of cross-attention mechanism in transformer with UAling)}\label{sec: casestudy}

\begin{figure*}[htbp]
    \centering
    \includegraphics[width=0.95\linewidth]{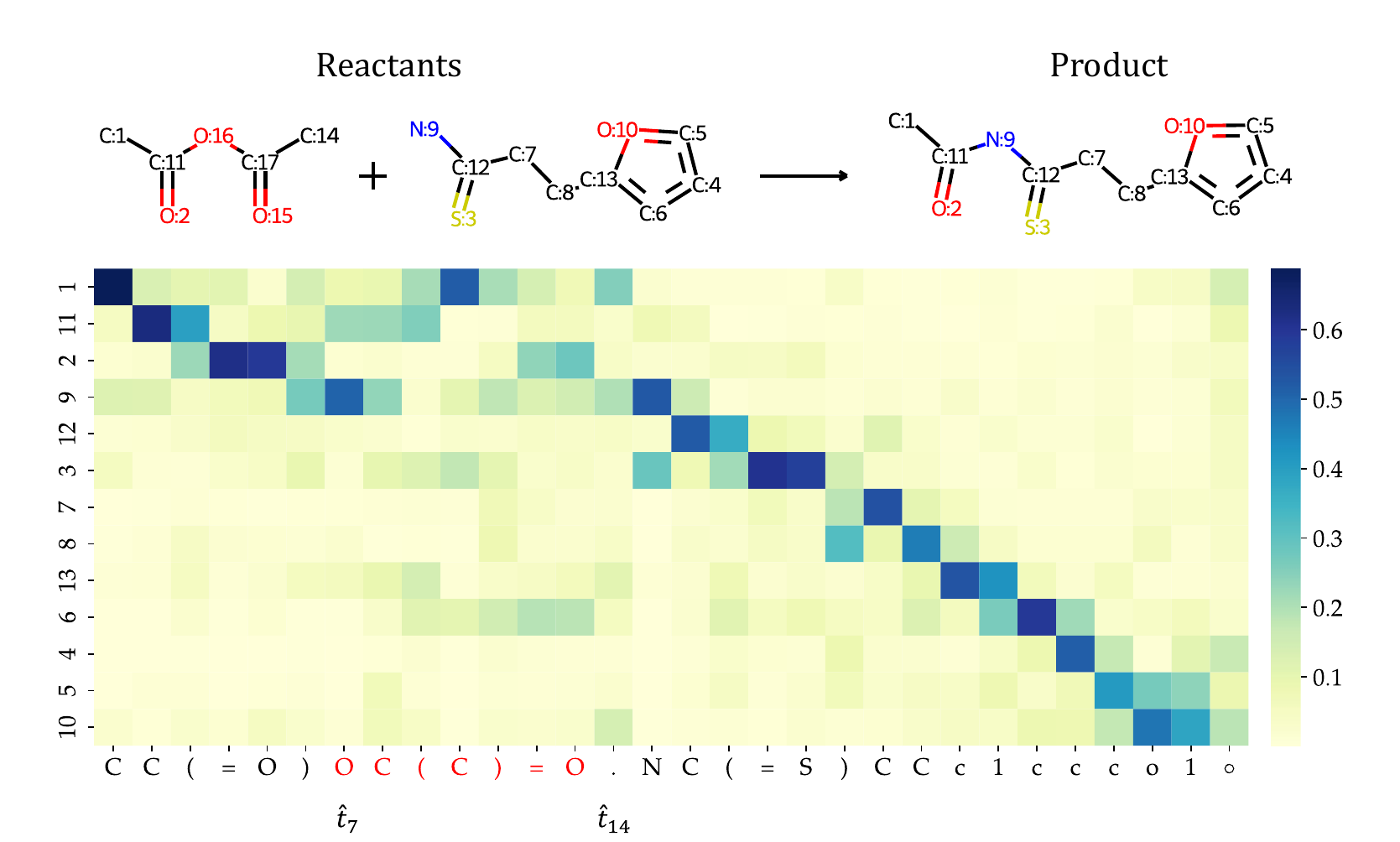}
    \caption{Visualization of cross-attention over order-aware node features and the predicted tokens. The number on the y-axis is the map number of atoms in the product. The reactants atoms that not appear in product is colored red in the x-axis. $\circ$ represents the end token.}
    \label{fig: case}
\end{figure*}
We randomly select a case from the dataset and showcase the cross-attention map in Fig.~\ref{fig: case}. The cross-attention map indicates the correlation between reactant tokens and nodes in the input product graph. This map is obtained by averaging the attention coefficient from each attention head. From the figure, it is evident that the predicted tokens successfully locate their corresponding atoms in the product, which contributes to the accurate prediction. The SMILES alignment can also be observed to assist the model in correctly identifying the reaction center. In accordance with the figure, the bond between atom \texttt{C:11} and \texttt{N:9} breaks during the transformation into reactants. Our model  effectively notices this occurrence and focus the attention of token $\hat{t}_7$ on the reaction centers \texttt{C:11} and \texttt{N:9}. This strategic focusing successfully guides the completion of the reactants, ensuring that the leaving group is correctly attached to the appropriate atoms. Additionally, we note that the attention coefficient at token $\hat{t}_{14}$ is concentrated on atoms \texttt{C:1} and \texttt{N:9}, which are the first atoms of each reactant molecule according to the given DFS order. This further indicates that our model is able to correctly identify the sites where the reaction occurs and accurately cleave the chemical bonds. Moreover, the attention of newly generated structures (i.e., tokens $\hat{t}_7$ to $\hat{t}_{13}$) is directed towards atoms \texttt{C:1}, \texttt{C:11}, and \texttt{O:2}, which correspond to the specific synthon they will attach to. This demonstrates that our model is able to generate appropriate functional groups based on the molecular structure information to form the reactants. All the aforementioned results illustrate that our proposed SMILES alignment method assists the model in comprehending molecular structural information and helps it to focus on learning chemical rules.

To further investigate the impact of the proposed SMILES alignment mechanism on model training, we visualize the cross-attention coefficients of different Transformer decoder layers. The visualization is provided in Supplementary Fig. 1 of Supplementary Information. From Supplementary Fig. 1, we can observe significant variations in the cross-attention across different layers. Additionally, the establishment of correspondence does not occur exclusively at certain layers, such as the first or last layer. This suggests that directly imposing supervised signals on the cross attention coefficient~\cite{wan2022retroformer, GTA} for SMILES alignment is not a wise approach, whether applied to all layers or only the last layer. It also verifies our statement in Sec.~\ref{sec: smalign} that using unsupervised methods for SMILES alignment does not affect the diversity of attention maps and further has a negative impact on model training and performance. This is why our unsupervised SMILES alignment mechanism achieves better results than supervised SMILES alignment.

\subsection{Case Study (Multi-step Retrosynthetic Pathway Planning)}

\begin{figure}[htbp]
    \centering
    \includegraphics[width=\linewidth]{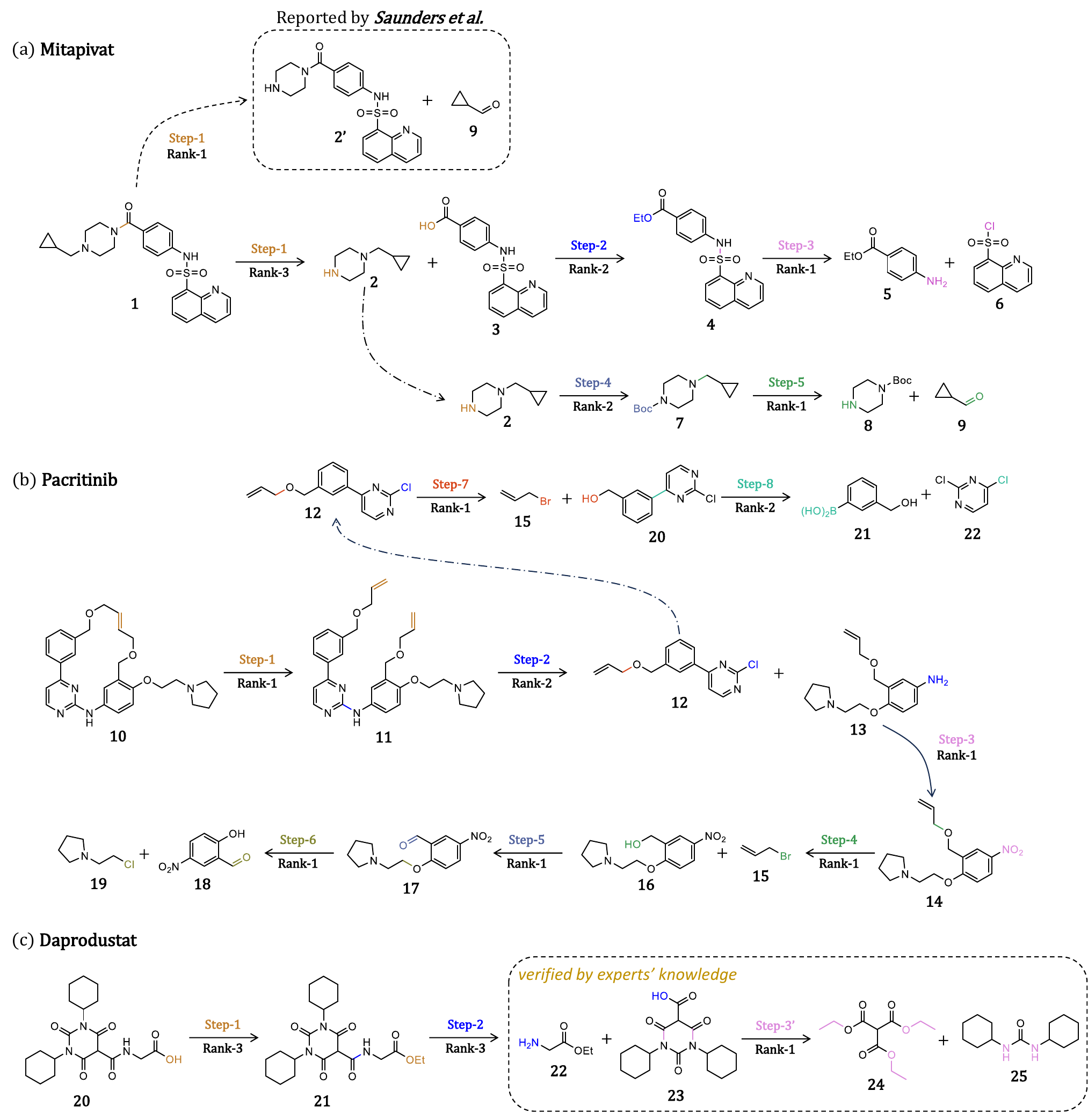}
    \caption{Multistep retrosynthesis predictions by our method. (a) Mitapivat (b) Pacritinib citrate (c) Daprodust. The reaction centers and leaving groups are highlighted in different colors. The pathway pf molecule (a) and (b) come from  literature, while the last one is verified by chemical experts.}
    \label{fig: case_path}
\end{figure}

To explore the suitability of our model for multi-step retrosynthetic pathway planning, we select three distinct molecules as targets for synthetic route design, and the synthesis routes are obtained through iterative calls to our UAlign model, which is trained with the USPTO-FULL dataset. The predicted pathways are summarized in Fig.~\ref{fig: case_path}.

The first case study involves Mitapivat, a compound approved for the treatment of hereditary hemolytic anemias in February 2022~\cite{doi:10.1177/20406207211066070}. Our model successfully predicted the five-step synthetic route reported in~\cite{benedetto2022fda}, with each step consistently ranked within the top-2 predictions. The first step entails an amide coupling reaction, which our model placed at rank 2, yielding the reactants 1-(cyclopropylmethyl)piperazine (compound \textbf{2}) and 4-(quinoline-8-sulfonamido)benzoic acid (compound \textbf{3}). Notably, at the initial step, our model also proposed an alternative synthesis method utilizing the Borch Reductive Amination, which was ranked at the first and is consistent with the synthetic route delineated by~\citeauthor{saunders2011therapeutic}. Subsequently, for the synthesis of 4-(quinoline-8-sulfonamido)benzoic acid, the model precisely executed a functional group protection strategy during the second step and accurately anticipated the subsequent formation of the sulfonamide, effectively deconstructing the target molecule into readily available precursors. For the synthesis of 1-(cyclopropylmethyl)piperazine, the model strategically protected the amine functional group with a tert-butyloxycarbonyl moiety at the outset and, in the ultimate step, prognosticated the N-alkylation reaction with a top-ranking accuracy. This example illustrates our model's capability to uncover diverse reaction centers in molecular retrosynthetic design and to generate plausible reactant combinations based on these insights.

The second case under scrutiny is Pacritinib, an orally bioavailable and isoform selective JAK-2 inhibitor for the treatment of patients with myelofibrosis, which received FDA approval on February 28, 2022~\cite{zhang2023synthesis}. As shown in Figure~\ref{fig: case_path}(b), our model successfully delineates a eight-step synthesis, as described in the literature~\cite{Pacrt}, tracing the synthetic pathway from commercially available 5-nitrosalicylaldehyde and 2,6-dichloropyrimidine to the final product. The initial step of the reverse synthesis is olefin metathesis, ranking the first in order of likelihood, followed by another rank-2 aromatic substitution of 4-(3-((allyloxy)methyl)phenyl)-2-chloropyrimidine (compound \textbf{12}) and 3-((allyloxy)methyl)-4-(2-(pyrrolidin-1-yl)ethoxy)aniline (compound \textbf{13}). Subsequently, synthesis of 4-(3-((allyloxy)methyl)phenyl)-2-chloropyrimidine was correctly identified via continuous allyl substitution and Suzuki cross-coupling reaction as the top and the second choices. The reverse synthesis of 3-((allyloxy)methyl)-4-(2-(pyrrolidin-1-yl)ethoxy)aniline was reduction of the nitro group, followed by another allyl substitution. In the final step, the model’s highest probability prediction was reduction of the aldehyde group, followed by a nucleophilic substitution. Despite the synthesis route involving a considerable number of steps and encompassing a variety of reaction types, our model successfully and accurately predicted each step within the top-2 choices. This accomplishment signifies the robustness and efficacy of our model in the context of retrosynthetic analysis.

The final case is Daprodustat, the first oral hypoxy-inducing factor prolyl hydroxylase inhibitor (HIF-PHI) for the treatment of renal anemia caused by chronic kidney disease (CKD)~\cite{HARA2015410}. This novel compound received approval for market release from the FDA on the 1st of February, 2023~\cite{zhang2023synthesis}. Our model predicted the three-step synthetic route. The first step reports the hydrolysis of ester at rank 3, which is aligned with the route provided by~\citeauthor{duffy2007preparation}. Although next two steps provided by our method do not exist in the literature, there are all explainable. The synthesis of ethyl (1,3-dicyclohexyl-2,4,6-trioxohexahydropyrimidine-5-carbonyl)glycinate (compound \textbf{21}) was identified via dehydration condensation of 1,3-dicyclohexyl-2,4,6-trioxohexahydropyrimidine-5-carboxylic acid (compound \textbf{23}) and ethyl glycinate (compound \textbf{22}) as the top choice, which avoided using toxic ethyl isocyanatoacetate reported in literature. In the final step, the model’s highest probability prediction was amidation of ester,  resulting in cost-effective and readily accessible starting materials. This case demonstrates the robust extrapolative capacity of our model, highlighting its potential to generate synthetic routes that surpass those documented in the literature.



\section{Limitations}
This work does not integrate much domain knowledge related to chemical reaction mechanisms in its design, which to some extent, compromises its interpretability. Similarly to most template-free methods, our work also faces challenges in generating diverse results. These aspects will be left for exploration in future works.

\section{Conclusion}\label{sec13}

We present {\modelname}, a novel graph-to-sequence pipeline that achieves state-of-the-art performance in the field of template-free methods. Our approach outperforms existing template-free and semi-template-based methods, while achieving comparable results to template-based methods. By utilizing a specially-designed graph neural network as the encoder, our model effectively leverages chemical and structural information from molecule graphs, resulting in powerful embedding for the decoder. Additionally, Our proposed unsupervised SMILES alignment mechanism facilitates the reuse of shared substructures between reactants and products for generation, allowing the model to prioritize chemical knowledge  even without complex data annotations. This significantly enhances the performance of the pipeline. In future work, we plan to further explore multi-step retrosynthesis planning using our {\modelname} as the single-step retrosynthesis prediction backbone.

\backmatter

\bmhead{Supplementary information}

We provide the details about our metrics, data preprocess and more visualization results in another pdf file as Supplementary Information.



\section*{Declarations}
\subsection{Funding}
This work was supported by the Shanghai Municipal Science and Technology Major Project (2021SHZDZX0102), the SJTU AI for Science platform, and the Fundamental Research Funds for the Central Universities.

\subsection{Competing interests}
The authors declare no competing interests.
\subsection{Data availability}
The Data of USPTO-FULL and USPTO-50K can be found in \url{https://github.com/Hanjun-Dai/GLN}. The Data of USPTO-MIT can be found in \url{https://github.com/wengong-jin/nips17-rexgen/tree/master/USPTO}. We also provide the raw data and our processed version in \url{https://github.com/zengkaipeng/UAlign}.
\subsection{Code availablity}
All the codes and checkpoints can be found in \url{https://github.com/zengkaipeng/UAlign}.
\subsection{Author contribution}
K.Z. was responsible for the code implementation, algorithm design, and the majority of the manuscript writing. X.Z. conducted the Ablation Study. Y.Z. and B.Y. managed multi-step Retrosynthetic Pathway Planning. F.N. handled all visualizations. The remaining authors contributed to the polishing of the article. Y.X., Y.J. and X.Y. supervised the research.
\subsection{Ethics approval and consent to participate}
Not applicable
\subsection{Materials availability}
Not applicable
\subsection{Consent for publication}
Not applicable

\bibliography{sn-bibliography,NeurIPS-2020-self-supervised-graph-transformer-on-large-scale-molecular-data-Bibtex,pmlr-v97-lee19d,reference}

\end{document}


\title[Article Title]{\normalsize{Supplementary material of}

{\ }

{\ }

\LARGE{
UAlign: Pushing the Limit of Template-free Retrosynthesis Prediction with Unsupervised SMILES Alignment}}


\author[1]{\fnm{Kaipeng} \sur{Zeng}}
\author[2]{\fnm{Bo} \sur{Yang}}
\author[1]{\fnm{Xin} \sur{Zhao}}
\author[1]{\fnm{Yu} \sur{Zhang}}
\author[3]{\fnm{Fan} \sur{Nie}}
\author[1]{\fnm{Xiaokang} \sur{Yang}}
\author*[1]{\fnm{Yaohui} \sur{Jin}}\email{jinyh@sjtu.edu.cn}
\author*[1]{\fnm{Yanyan} \sur{Xu}}\email{yanyanxu@sjtu.edu.cn}
\affil*[1]{\orgdiv{MoE Key Laboratory of Artificial Intelligence, AI Institute}, \orgname{Shanghai Jiao Tong University}, \orgaddress{\city{Shanghai}, \postcode{200240}, \state{Shanghai}, \country{China}}}

\affil[2]{\orgdiv{Frontiers Science Center for Transformative Molecules (FSCTM), Zhangjiang Institute for Advanced Study}, \orgname{Shanghai Jiao Tong University}, \orgaddress{\city{Shanghai}, \postcode{200240}, \state{Shanghai}, \country{China}}}
\affil[3]{\orgdiv{Department of Computer Science and Engineering}, \orgname{Shanghai Jiao Tong University}, \orgaddress{\city{Shanghai}, \postcode{200240}, \state{Shanghai}, \country{China}}}

\maketitle

\section{Notations}
Supplementary Table~\ref{tab:notations} lists the notations used to facilitate reading.
\begin{sidewaystable}[htbp]
\centering
\captionsetup{name={Supplementary Table}}
\caption{Notations for facilitating reading.}
\label{tab:notations}
\begin{tabular}{c|l}
\hline
\textbf{Notation}  & \makecell[l]{\textbf{Description}} \\
\hline
$G=(V,E)$& a graph with nodes $V$ and edges $E$\\
$rank(a, O)$& the position of atom $a$ in DFS order $O$ \\
$root(G,O)$& the atom (node) of G with the minal rank in $O$\\
$Smiles(G,O)$& The SMILES of molecule graph $G$ generated via DFS order $O$\\
\hline
$M=(V_M,E_M)$ & a molecule graphs\\
$M'=(V_{M'}, E_{M'})$ & a molecule graph\\
$\tilde{M}=(V_M \cup V_{M'}, E_M \cup E_{M'})$& a molecule graph consist of two isolated molecules $M$ and $M'$\\
$O_M$ & a DFS order of molecule graph $M$\\
$O_{\tilde{M}}$& a DFS order for molecule graph $\tilde{M}$\\
\hline
$\mathcal{R}$& the set of all reactant molecule graphs\\
$P=(V_P, E_P)$       &  the product molecule graph \\
$R=(V_R, E_R)$ & a single reactant molecule graph\\
$\mathcal{D}(V_P), \mathcal{D}(V_R)$ & the set of all possible DFS order for product molecule graph / a single reactant molecule graph\\
$O_P, O_R$ & a DFS order of product, a DFS order of reactant molecule graphs\\
$CO(R,O_P)$ & the DFS order of reactant $R$ that has a nearly consistent atomic appearance sequence with $O_P$ \\
$OPSmiles(\mathcal{R}, O_P)$ & the order-preserving reactants SMILES of $\mathcal{R}$ based on DFS order $O_P$\\
\hline
$h^{(k)}_u$ & the node feature of node $u$ at $k$-th iteration of message passing.\\
$e^{(k)}_{u,v}$ & the edge feature of edge between node $u$ and $v$ at $k$-th iteration of message passing.\\
$H$&the output of the encoder, i.e. the encoded node features\\
\hline
$Z$ & the input token embedding for Transformer decoder\\
$\hat{H}$& the order-aware node features\\
$\hat{Z}$ & the output token embedding of Transformer decoder\\
\hline
$\hat{T}$& the predicted token logits list \\
$\hat{t}_i$& the $i$-th token logits in the predicted token list\\
$T$& the ground truth token list\\
$t_i$& the $i$-th ground truth token\\
\hline
\end{tabular}
\end{sidewaystable}

\section{Dataset}\label{app: data}

The open benchmark datasets are presented as follows.

\textbf{USPTO-50K.} The USPTO-50K is of high quality, we only remove the reactions whose products only contain a single atom from the dataset.

\textbf{USPTO-FULL.} The raw data of USPTO-FULL provided by \citeauthor{gln} has error annotations and other problems. Thus we perform the following steps to clean up the data.
\begin{itemize}
    \item remove the reactions where different atoms share the same atom map number.
    \item remove the reactions whose product is consist of a single atom.
    \item remove the reactions with invalid SMILES. The empty SMILES is considered as invalid SMILES too.
    \item remove the reactions where product contains atoms that do not appear in reactants.
    \item remove the reactants that all atoms do not appear in the product.
\end{itemize}

\textbf{USPTO-MIT.} The raw data of USPTO-MIT provided by~\cite{jin2017predicting}. The original dataset contains reactions that have multiple product molecules. And the reagents are put together with reactions. Thus we perform the following steps to make the dataset more suitable for retrosynthesis prediction task.
\begin{itemize}
    \item remove the reactions whose product is consist of a single atom
    \item remove the reagents from the reactants. We consider the molecules whose atoms have no intersection with the product as reagents.
    \item remove the reactions with multiple product molecules.
\end{itemize}

The statistical information of the datasets used in this work is summarized in Supplementary Table~\ref{tab:dataset}.

\begin{table}[htbp]
\captionsetup{name={Supplementary Table}}
    \caption{\textbf{Summary of datasets used in this paper. }\#Train/\#Valid/\#Test denotes the number of samples in the training/validation/test set, respectively. \#Total is the sum of \#Train, \#Valid and \#Test. }\label{tab:dataset}
    \centering
     \begin{tabular}{ccccc}
     \toprule
         \textbf{Dataset} &
         \textbf{\#Train} &
         \textbf{\#Valid} &
         \textbf{\#Test} &
         \textbf{\#Total} \\
         \hline
        \textbf{USPTO-50K} & 40,006
         & 5,001
         & 5,007
         & 50,016
         \\
         \textbf{USPTO-MIT} & 395,498 & 29,076 & 38,648 & 453,222\\
          \textbf{USPTO-FULL}
         & 768,679
         & 96,076
         & 96,015
         & 960,770\\
         \bottomrule
    \end{tabular}
\end{table}

\section{Evaluation Metrics}\label{sec: eval metric}
\textbf{Top-$k$ Accuracy.} We use the conventional top-$k$ accuracy to evaluate the performance of model. A prediction result is considered as correct if and only if all the reactants are correctly predicted.

\textbf{Top-$k$ SMILES Validity.} As SMILES is considered as correct when it can be identified by RDKit~\cite{landrum2013rdkit}. The top-$k$ SMILES validity is calculated as $\frac{1}{N\times k}\sum_{i=1}^N\sum_{j=1}^k \mathbbm{1}(\mathrm{SMILES\ is\ valid})$, where $N$ is the number of evaluated samples.

\textbf{Top-$k$ round-trip Accuracy.} There may be multiple methods to synthesize a given product. Evaluating the model's performance solely based on top-k accuracy may lead to biased results. Therefore, in addition to top-k accuracy, we also employ top-k round-trip accuracy as an extra metric to evaluate the model's performance. Top-k round-trip accuracy measures the percentage of predicted reactants that can undergo a reaction and yield the given product. To align with our baseline, we use the Molecule Transformer~\cite{schwaller2019molecular} as the forward reaction prediction model. The calculation of top-k round-trip accuracy follows the same approach as described in~\cite{wan2022retroformer}.

\section{Visualization of the Cross-Attention Coefficients}
We visualize the cross-attention maps across different transformer decoder layers using the checkpoint trained on the USPTO-50K dataset. The outcomes are depicted in Supplementary  Fig.~\ref{fig: sup_crs}.
\begin{figure}[htbp]
    \centering
    \includegraphics[width=\linewidth]{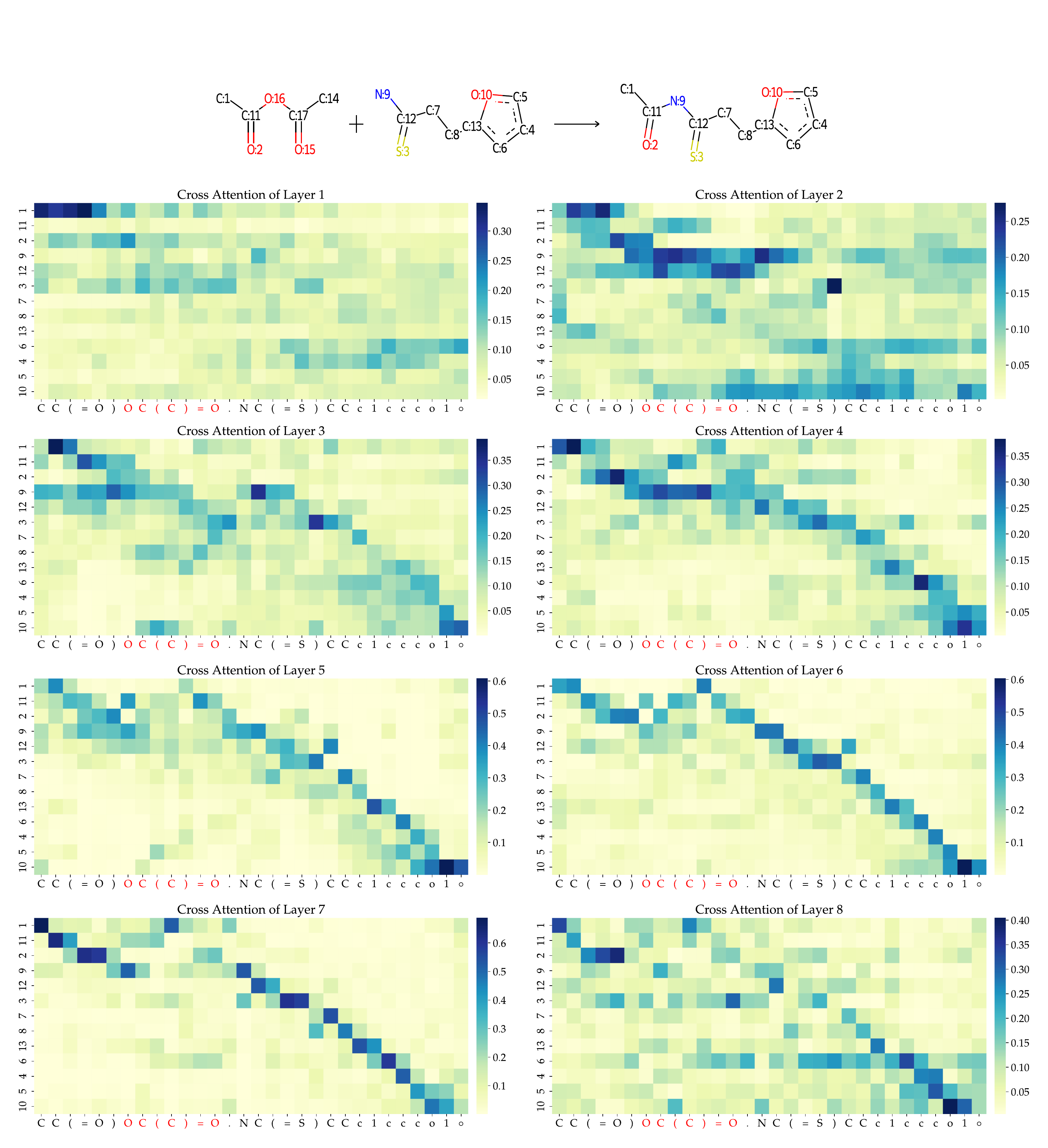}
    \captionsetup{name={Supplementary Fig}}
    \caption{Visualization of cross attention over product atoms and predicted tokens of different Transformer decoder layers. The number on the y-axis is the map number of atoms in the product. The reactants atoms that not appear in product is colored red in the x-axis. $\circ$ represents the end token.}
    \label{fig: sup_crs}
\end{figure}

\section{Implementation details} \label{impdetail}
\subsection{Generation of Order-Preserving Reactant SMILES}\label{opsmildetail}
Given a traversal order $O_P$ of product $P$, there might be multiple DFS orders of a reactant molecule that satisfy the condition to be the $O_P$-corresponding order. This is because reactants might contain atoms and substructures that are not present in the final product. Also, directly solving this problems is very difficult. Thus, we use an approximate  algorithm, which is shown in Algorithm~\ref{alg: CO} to generate the corresponding order defined in Eq. 2 given product molecule $P$, reactant molecule $R$ and a DFS order $O_P$ of $P$, where $am(i)$ represents the atom map number of atom $i$.

\begin{algorithm}[htbp]
   \caption{Get Corresponding Order}
   \label{alg: CO}
    \begin{algorithmic}
       \State {\bfseries Input:} product graph $P=(V_P, E_P)$, DFS order of product $O_P$, reactant $R=(V_R, E_R)$.
       \Function{get\_rank}{$O_P$: DFS order of $P$, $x$: atom, $R$: reactant, $AMRank$: a dict, $Vis$: a set}
       \State Add $x$ into $Vis$.
       \State Initialize $min\_son$ as $\infty$.
       \For{$y\in \mathrm{GET\_NEIGHBOR(x, R)}$}
            \If{$y\notin Vis$}
            \State GET\_RANK($O_P$, y, $R$, $AMRank$, $Vis$).
            \State $min\_son = \min(min\_son, AMRank[am(y)])$.
            \EndIf
       \EndFor
       \If{$x \in O_P$}
        \State $AMRank[am(x)]=rank(x, O_P)$.
        \Else
        \State $AMRank[am(x)]= min\_son$.
        \EndIf
       \EndFunction
       \Function{get\_CO}{$CO$: a list, $x$: atom, $R$: reactant, $AMRank$: a dict, $Vis$: a set}
        \State Add $x$ into $Vis$.
        \State Append $x$ to $CO$.
        \State Initialize $N$ as an empty list.
        \For{$y\in \mathrm{GET\_NEIGHBOR(x, R)}$}
            \If{$y\notin Vis$}
                \State Append $y$ to $N$.
            \EndIf
       \EndFor
       \State Sort $N$ in ascending order according to $AMRank[am(x)], x\in N$.
       \For{$y\in N$}
        \State GET\_CO($CO$, $y$, $R$, $AMRank$, $Vis$).
       \EndFor
       \EndFunction
       \State Initialize $Vis$ as an empty set.
       \State Initialize $CO$ as an empty list.
       \State Initialize $AMRank$ as a empty dict.
       \State Set $root$ as the atom $x\in V_R$ that $am(x) = am(root(P, O_P))$.
       \State GET\_RANK($O_P$, $root$, $R$, $AMRank$, $Vis$)
       \State Reset $Vis$ as an empty set.
       \State GET\_CO($CO$, $root$, $R$, $AMRank$, $Vis$).
       \State {\bfseries Output:} $CO(R, O_P)=CO$
    \end{algorithmic}
\end{algorithm}

\subsection{Model Implementation details (Training and Inference)}
We implement our model based on \texttt{Pytorch 1.13}~\cite{paszke2019pytorch} and \texttt{torch\_geometric 2.2.0}~\cite{Fey/Lenssen/2019pyg}.  For model for USPTO-50K, we set the hidden size as 512, encoder layer as 8, decoder layer as 8 and the number of attention heads as 8. For model for USPTO-FULL, we set the hidden size as 768, encoder layer as 8, decoder layer as 8 and the number of attention heads as 12. The dropout ratio for both model is set  as 0.1. The highest learning rate of each model as set as 1.25e-4. We slowly increase our learning rate to the highest in the first few epochs and slowly decrease it using exponential decay. The model is trained with Adam optimizer~\cite{kingma2014adam}. The maximum number of training epochs for both stages is established at 300, and an early stopping strategy is implemented, which terminates the training process if the model fails to demonstrate performance improvement on the validation set for ten consecutive epochs.

It is important to note that a prescribed DFS order may implicitly capture the distribution of reactant sub-groups within the product, which could inadvertently result in information leakage.  To mitigate this issue, we employ a DFS order during the inference process that generates a canonical SMILES representation for the product molecule, ensuring a standardized input that precludes information leakage.

\section{Initial Node Features and Edge Features}\label{init_feat}
We use the atom and bond descriptors provided by \texttt{Open Graph Benchmark}~\cite{hu2020open}. There are nine atom descriptors including atomic number, chirality, formal charge and other properties that can be calculated by RDKit~\cite{landrum2013rdkit}. And three bond descriptors are used,  containing bond type, bond stereochemistry as well as whether the bond is conjugated. All the descriptors provide integer outputs. Thus we set up learnable embedding tables for each descriptor.  The initial node features and edge features are then obtained by summing up the embedding corresponding to the output of each descriptor.
\bibliography{supplement,sn-bibliography}